\documentstyle[aas2pp4,psfig]{article}

\newcommand{\lesim}{\,\raisebox{-0.4ex}{$\stackrel{<}{\scriptstyle\sim}$}\,}
\newcommand{\PSR}{PSR~B1259$-$63}
\newcommand{\Rstar}{R_\ast}

\newcommand{\Modot}{{\rm M}_\odot}
\newcommand{\Rodot}{{\rm R}_\odot}
\newcommand{\peri}{{\cal T}}

\newcommand{\rms}{{\rm s}}
\newcommand{\rmp}{{\rm p}}
\newcommand{\rmw}{{\rm w}}
\newcommand{\rmd}{{\rm d}}

\newcommand{\rs}{r_\rms}
\newcommand{\rp}{r_\rmp}
\newcommand{\ns}{n_\rms}
\newcommand{\np}{n_\rmp}

\newcommand{\Bs}{B_\rms}
\newcommand{\Bp}{B_\rmp}

\newcommand{\vw}{v_\rmw}
\newcommand{\vs}{v_\rms}
\newcommand{\vdo}{v_{\rmd 0}}
\newcommand{\nd}{n_\rmd}
\newcommand{\ndo}{n_{\rmd 0}}
\newcommand{\nwo}{n_{\rmw 0}}
\newcommand{\sigmad}{\sigma_\rmd}

\newcommand{\dotMd}{\dot{M}_\rmd}
\newcommand{\dotMw}{\dot{M}_\rmw}

\newcommand{\Edot}{\dot{E}_\rmp}
\begin{document}

\title{ORIGIN OF THE TRANSIENT UNPULSED RADIO EMISSION FROM THE \PSR\ 
 BINARY SYSTEM}
\author{Lewis Ball$^1$, Andrew Melatos$^{2}$, Simon Johnston$^1$
\& Olaf Skj\ae raasen$^{1,3}$}
\affil{$^1$ Research Centre for Theoretical Astrophysics, University of Sydney,
NSW 2006, Australia; ball@physics.usyd.edu.au}
\affil{$^2$ Miller Fellow, Department of Astronomy, 601 Campbell Hall,
University of California, Berkeley CA 94720 USA}
\affil{$^3$ Institute of Theoretical Astrophysics, University
of Oslo, P.O. Box 1029 Blindern, N-0315 Oslo, Norway}

\vspace{\baselineskip}
\centerline{Astrophysical Journal Letters, in press}
\centerline{Original: 4 May 1998. Revised: 29 June 1998, 10 December 1998}

\keywords{binaries: eclipsing -- circumstellar matter --
stars: emission-line, Be -- radio continuum: stars
-- pulsars: individual: \PSR\ }

\begin{abstract}
We discuss the interpretation of transient, unpulsed radio 
emission detected from the unique pulsar/Be-star binary system 
PSR B1259$-$63.
Extensive monitoring of the 1994 and 1997 periastron passages 
has shown that the source flares over a 100-day interval around
periastron, varying on time-scales as short as a day and peaking
at 60 mJy ($\sim 100$ times the apastron flux density)
at 1.4 GHz.
Interpreting the emission as synchrotron radiation,
we show that (i) the observed variations in flux density
are too large to be caused by the shock interaction between
the pulsar wind and an isotropic, radiatively driven,
Be-star wind, and (ii) the radio emitting electrons do not originate
from the pulsar wind.
We argue instead that the radio electrons originate from
the circumstellar disk of the Be star and are accelerated at 
two epochs, one before and one after periastron, when the pulsar 
passes through the disk.
A simple model incorporating two epochs of impulsive acceleration
followed by synchrotron cooling
reproduces the essential features of the radio light curve
and spectrum and is consistent with the
system geometry inferred from pulsed radio data.

\end{abstract}

\section{Introduction}

The binary pulsar \PSR\ is in a highly eccentric
($e\sim 0.87$), 3.5-yr orbit around a 10th magnitude B2e star,
SS 2883, of radius $\Rstar\sim 6\;\Rodot$
(Johnston et al.\ 1992, 1994).
The system is unique because \PSR\ is the only radio pulsar
orbiting a main-sequence companion with a circumstellar disk.
Observations of the H$\alpha$ emission line show that optically emitting
material in the Be-star disk extends to at least $20\;\Rstar$
(Johnston et al.\ 1994),
and pulsar timing data suggest that the disk is steeply inclined
to the plane of the binary orbit (Wex et al.\ 1998).
The pulsar's orbital radius at periastron is $23\;\Rstar$.
Melatos, Johnston \& Melrose (1995) proposed a wind-disk
model of the system to explain orbital variations of the
flux density, polarisation,
rotation measure and dispersion measure of the pulsed emission
near periastron (Johnston et al.\ 1996).
In the model, the eclipse of the pulsed emission between $\peri -20$
and $\peri + 16$ (where $\peri$ denotes the epoch of periastron)
is attributed to free-free absorption in the disk.

X-ray observations of \PSR\ reveal significant (unpulsed) emission
throughout the orbit, with the flux density at periastron
$\sim 10$ times higher than at apastron
(Hirayama et al.\ 1996).
Modeling suggests that the X-rays are synchrotron radiation
from relativistic pulsar wind leptons accelerated at the
shock between the pulsar wind and 
Be-star outflow (Tavani \& Arons 1997).

We discuss the origin of the transient, unpulsed radio emission 
detected from \PSR\ from $\peri -22$ until $\peri +100$
(Johnston et al.\ 1999). 
The radio emission, like the X-rays, is likely
to be synchrotron radiation produced
by the shock interaction between the
pulsar wind and Be-star outflow.
However, we argue that the interaction between the pulsar wind
and an isotropic, radiatively driven, Be-star wind cannot explain
the magnitude and rapidity of the variations in the radio light curve.
Furthermore, we show that the radio emitting electrons, unlike the X-ray
electrons, do not originate from the pulsar wind.
We therefore suggest that the unpulsed radio emission
results from the acceleration of electrons in the Be-star disk.
The pulsar--disk encounter is likely to result in a strongly
time-dependent, magnetized, hydrodynamical interaction
forming a cometary pulsar wind nebula surrounded by
a shock front where acceleration of particles is followed by a
combination of adiabatic, synchrotron and inverse Compton cooling.
However, the data so far obtained from this system are insufficient
to constrain a model of the physics at this level of detail.
Here we propose a simple model for the radio light curves
which reproduces the main features of the data and
identifies the essential physics.

\section{Observations}

Two periastron passages of \PSR\ have been observed in detail,
the first on 1994 Jan 9 and the second on 1997 May 29.
Unpulsed radio emission was detected at
frequencies between 0.84 and 8.4~GHz, 
as described by Johnston et al.\ (1999).
Figure 4 of that paper shows the radio light curve at five
frequencies during the 1997 periastron passage
(the observed 1.4~GHz light curve is reproduced here
in Figure 1b).
Emission appeared suddenly at $\peri-22$;
prior to this the 3-$\sigma$ upper limit on the
unpulsed emission is $\lesim 0.5$ mJy at
frequencies between 1.4 and 8.4~GHz.
For the next few days, rapid variations were superimposed
on an overall increase in the flux density $S$ at all frequencies,
peaking at $S \sim 30$~mJy at 1.4~GHz near $\peri -8$.
There followed a relatively slow decline
before the flux density again began increasing around $\peri+8$,
peaking a second time near
$\peri+18$, with $S\sim 60$~mJy at 1.4~GHz.
The flux density then declined steeply until around $\peri+30$,
when the rate of decay slowed to roughly that observed after the
first peak near $\peri-8$.
This slower decline continued until the last observations
on $\peri+47$.
The 1994 data are similar
and showed detectable emission until $\peri+100$.

At most times the radio frequency spectrum is well fitted by a
power law $S\propto \nu^{-\alpha}$ of spectral index 
$\alpha \approx 0.5$, consistent with synchrotron radiation
from relativistic electrons.
Importantly, there is evidence in both the 1994 and 1997 data
that the spectrum steepens after periastron
(Figure 5 of Johnston et al.\ 1999).
There is no compelling evidence for free-free
absorption at any epoch between $\peri-22$ and $\peri+47$
(Johnston et al.\ 1999),
even at periastron when the pulsed emission is totally
eclipsed by the Be-star disk.

\section{Origin of the radiating electrons}

The relativistic pulsar wind is enclosed within a bubble
as it ploughs through the material around the Be star.
Momentum flux balance between the pulsar wind 
and Be-star outflow determines the size and shape of the 
bubble, and particles are shock accelerated at the boundary
between the two flows (Melatos, Johnston \& Melrose 1995).

The Be-star outflow consists of a radiation-driven
polar wind and a more dense equatorial disk
responsible for the bulk of the mass loss
(Dougherty 1994).
Both components are
often characterised as simple power laws in radial distance $\rs$,
with number density and speed given by
$\ns= n_0 (\Rstar / \rs)^\xi$
and $\vs =v_0 (\rs / \Rstar)^{\xi-2}$,
where $\xi=2$ corresponds to a constant speed
and $\xi=4$ corresponds to constant ram pressure.
If the outflow is highly conducting so that the magnetic field is
frozen into the plasma, then simple models imply
$B_\rms\propto \rs^{1-\xi}$ for $2\leq\xi\leq 3$.
The polar wind is generally thought to
reach a terminal speed $\vw \sim 1000 \; {\rm km \, s^{-1}}$
within a few stellar radii (e.g.\ Friend \& McGregor 1984);
we therefore approximate it as a constant speed outflow with
$\xi_\rmw=2$.
In contrast, the disk density falls off rapidly with
radius, as rotating material gradually
accelerates outward; one typically has $\xi_\rmd\lesim 4$,
$\vdo\sim 5\; {\rm km \, s^{-1}}$
and a half opening angle of $\sigmad\sim 5-15\arcdeg$
(Waters, Cot\'e \& Lamers 1987).
The disk and wind number densities are related by
$\ndo/\nwo=(\vw/\vdo) (\dotMd/\dotMw)(1-\sin\sigmad)/\sin\sigmad$.
For $\sigmad=10\arcdeg$ this implies a density ratio of
$\ndo/\nwo\sim 10^3 (\dotMd/\dotMw)$.
Typical values of $\dot M \sim 2 \times 10^{-8}\;\Modot\,{\rm yr^{-1}}$
and $\dotMd/\dotMw \sim 10$ yield
$\ndo\sim 4\times 10^{18}\;{\rm m^{-3}}$.

\subsection{Flux scaling for polar wind interaction}

If the spin-down luminosity of the pulsar,
$\Edot$, goes entirely into the relativistic wind 
then the ram pressure
at a distance $\rp$ from the pulsar is
$\Edot/(4\pi c \rp^{2})$.
The ram pressure of the Be-star polar wind is 
\linebreak
$m_{p} \ns(\rs) \vs(\rs)^{2}$.
The contact discontinuity between the two flows
has its apex at the point
where the ram pressures balance,
provided the magnetic pressures can be ignored.
This point is denoted by $\rp=D-\rs=L$ where $D$ is the
pulsar-Be star separation, $L/D=\sqrt{\eta}/(1+\sqrt{\eta})$
and $\eta=\Edot/(\dotMw c \vw)$
is the ratio of the momentum flux in the pulsar wind
to that in the Be-star polar wind.
The ratio $L/D$ is constant over the pulsar orbit,
although $D$ and $L$ both vary.

Now suppose that the unpulsed radio flux
is associated with the shock interaction between the
pulsar wind and Be-star polar wind.
Synchrotron theory implies that the flux density scales as
\begin{equation}
S\propto n \, l^3 B^{\alpha+1}\nu^{-\alpha}
\label{eq:synch}
\end{equation}
where $n$ is the number density of radiating electrons,
$l$ is the source size and $B$ is the magnetic field.
Assuming $l\propto L$ and that the other source parameters
scale as their values at the contact surface apex, 
one obtains
$l\propto D$, $n\propto D^{-2}$ and $B\propto D^{-1}$,
implying $S\propto D^{-1/2}$ for $\alpha=0.5$
(irrespective of whether the radiating electrons
originate from the pulsar wind or Be-star polar wind).
Since $D$ varies by a factor of $\sim 15$ around the orbit,
this argument implies that $S$ should be
$\sim 4$ times higher at periastron than at apastron
--- consistent with the sparse X-ray data,
but far smaller than the 1.4~GHz radio variation
from $S \lesim 0.5\;{\rm mJy}$ at apastron to 
$S \sim 60 \;{\rm mJy}$ near periastron.
Furthermore, in the interval near periastron,
the above scaling yields 
$S(\peri-27) \sim S(\peri)/\sqrt{3} \sim 20$~mJy,
yet at $\peri -27$ there is no detectable unpulsed
radio emission at any frequency.
We conclude that the emission
cannot be attributed solely to the interaction between 
the pulsar wind and Be-star polar wind.

\subsection{Disk or pulsar electrons? --- \\Lorentz factor}

The pulsar orbit is eccentric and steeply inclined
with respect to the Be-star disk, so the pulsar wind
interacts strongly with the disk near periastron,
potentially generating unpulsed radio emission.
We now investigate whether
the radiating electrons in this scenario
originate from the pulsar wind or the disk.

The magnetic field in the emission region is the sum
of the contribution from the pulsar,
$\Bp(L)={1.5\times 10^{-3} \Rstar/L}$~tesla
(estimated from $P$, $\dot P$ and the assumption
$\Bp \propto \rp^{-1}$ in the pulsar wind),
and the contribution from the Be star.
The field in the Be-star polar wind satisfies
$\Bs(D-L)\lesim{10^{-2} \Rstar/(D-L)}$~tesla
(see Barker 1987 for an upper limit on surface fields
of Be stars), while estimates of the field in the
disk are so uncertain as to be of little value.
For $L\lesim D/2$, one finds $B\sim 10^{-4}$~T at
$D\sim 50\Rstar$ when the radio peaks occur,
and hence $\Bp(L)\sim \Bs(D-L)$.
This value of $B$ is consistent with that obtained
independently from synchrotron equipartition arguments,
which yield $B\sim 10^{-4}$~T
for $S(1\;{\rm GHz})\sim 50\;{\rm mJy}$ and $\alpha=0.5$,
assuming a source distance of 1.5~kpc,
a spherical source of radius $25\;\Rstar$.

The Lorentz factor required to produce synchrotron emission at 1~GHz
in such a field is $\gamma\sim 50(\rp/50\Rstar)^{1/2}$ ---
five orders of magnitude lower than the likely Lorentz factor
$\gamma\sim 10^6$ of unaccelerated leptons in the pulsar wind
(Kennel \& Coroniti 1984; Arons \& Gallant 1994).
We conclude that the pulsar-wind electrons
are too energetic to be
responsible for the unpulsed radio emission observed.
On the other hand,
non-relativistic electrons from the Be-star disk and/or
polar wind could certainly be accelerated to the
relatively modest energy required at shock(s) produced
by the interaction between the pulsar wind and Be-star disk
(e.g.\ Kirk 1994)
as they are in binaries comprising early-type stars
(Eichler \& Usov 1993).

\subsection{Disk or pulsar electrons? --- \\Number density}

The number density of radiating electrons 
can be estimated straightforwardly for an assumed source size and
magnetic field,
since from synchrotron theory $n \propto S \, l^{-3} B^{-3/2}$.
The source size $l$ is at least the standoff
distance $L$, otherwise synchrotron self-absorption
would cause a detectable low-frequency turnover
below 1~GHz (Johnston et al.\ 1999).
If the emission results from the interaction between the
pulsar wind and Be-star disk, then 
the source is likely to be some kind of sheath
around the contact discontinuity, extending downstream
in a cometary tail.
A spherical sheath with
thickness $x\,L$, where $x$ is a numerical factor less than 1,
yields $n \sim x^{-1}\,10^{10}\;{\rm m^{-3}}$ 
for the parameters used above.

The number density of the pulsar wind can be estimated straightforwardly
from the spindown luminosity $\Edot=8.3\times 10^{28}$~W
and wind Lorentz factor $\gamma\sim 10^6$,
yielding
$\np(\rp) = {\Edot/ 4\pi \gamma m_e c^3 \rp^2}$
and hence $\np(L)\sim 2\times 10^4\;{\rm m^{-3}}$ for $L\sim 25\Rstar$.
This is at least 5 orders of magnitude smaller than the estimated
number density required to produce the observed radio flux density.
While the estimates clearly allow considerable latitude,
and a shock may compress the wind by a factor of order 10,
it is very difficult to argue away this discrepancy.
We conclude that the pulsar-wind electrons
are too few in number
to be responsible for the unpulsed radio emission observed.
On the other hand, even if the number density of the Be-star disk
falls off as steeply as $\nd\propto \rs^{-4}$,
one finds $\nd(25 \Rstar)\sim 10^{13}\;{\rm m^{-3}}$,
and acceleration of just $10^{-3}$ of the electrons in the
disk can account for the observed emission.

\section{Model of the radio light curve \\and spectrum}

Johnston et al.\ (1999) and Wex et al.\ (1998)
have argued that the Be-star disk is inclined with respect
to the orbital plane in such a way that the pulsar passes
through the disk twice in each orbit, once before periastron
(crossing the disk around $\peri-18$ at $D_1\approx 40\Rstar$)
and once afterwards ($\peri+13$ at $D_2\approx 33\Rstar$).
We postulate that the radiating electrons are created during
time intervals $T_1$ and $T_2$ at the two disk crossings,
when the shock interaction between the pulsar wind and the disk
(and hence electron acceleration) is at its peak,
and nowhere else.
$T_1$ and $T_2$ are somewhat greater than the orbital
crossing time of the disk 
($\sim 3\,{\rm days}$ with $\sigmad=5\arcdeg$
according to Johnston et al.\ 1999) 
because the pulsar wind interacts appreciably with the
disk surface for a few days before the pulsar enters
and after it leaves.

The two resulting electron populations are assumed to 
evolve with time purely under the action of synchrotron losses.
The loss times at the two crossings are given in terms of
constants $b_{1,2}=C (B_{1,2}^2/2\mu_0)$,
where $B_{1,2}$ is the local magnetic field,
$C=32\pi r_e^2/(9 m_e^2 c^3)$,
and $r_e$ is the classical electron radius.
The number densities per unit energy $N_{1,2}(\varepsilon,t)$ 
then obey
\begin{equation}
\frac{\partial N_{1,2}(\varepsilon,t)}{\partial t}
= b_{1,2} \frac{\partial}{\partial \varepsilon}
[ \varepsilon^2 N_{1,2}(\varepsilon,t) ]
+ Q_{1,2}(\varepsilon,t),
\label{eq:diffeqn}
\end{equation}
where $Q_{1,2}(\varepsilon,t)$, the injection rate
at each disk crossing, is a power law in energy
during the times $T_1$ and $T_2$ and zero otherwise.
Note that the evolution of the accelerated electrons
depends only on conditions at the acceleration site;
it is unrelated to the subsequent orbital position of
the pulsar.
Emission from the first disk crossing persists as the pulsar
rounds the Be star (behind the disk) and is still present 
when the pulsar re-enters the disk after periastron,
so the observed emission is the sum of that from the
two electron populations.

The lack of evidence of a low-frequency spectral turnover implies
that there is little absorption of the unpulsed radio emission
(Johnston et al.\ 1999),
even when the pulsar is at periastron
and is obscured behind the dense Be-star disk.
This strongly suggests that the radio-emitting electrons are
left behind by the pulsar in a ram-pressure-confined bubble
at the point where the pulsar crosses the disk, and that
therefore their radio emission is not obscured by the bulk
of the disk (Ball et al. 1998). 
We therefore ignore propagation effects on the radio emission.

The proposed model does not include the effects of
adiabatic expansion in the evolution of $N_{1,2}(\varepsilon,t)$.
Melatos, Johnston \& Melrose (1995) showed that when the pulsar is
deep within the Be-star disk, the momentum flux in the pulsar wind
is insufficient to disrupt the disk and so the pulsar wind
is contained within a bubble.
It follows that in a model which assumes that the bubble of
radiating electrons is spherical, the source does not expand and
so adiabatic losses do not occur.
Electrons accelerated during each disk
crossing are advected away from the shock apex in a
cometary backflow (Melatos, Johnston \& Melrose 1995),
but the observed rate of decline of the unpulsed radio emission
indicates that the resulting adiabatic losses must occur on a
timescale of a week or longer.
In this Letter we show that a model involving only injection
during the pulsar disk crossings and synchrotron losses is in
qualitative agreement with the observations.

Inverse Compton losses may be important in this system
(Tavani \& Arons 1997; Kirk, Ball \& Skj\ae raasen 1999).
At the energies of the radio-emitting electrons these losses
have the same form as synchrotron losses, and so the evolution of the
electron spectrum will still be described by an equation of the
form of (\ref{eq:diffeqn}).
Outside of the disk the energy density of target photons from
the Be star is comparable to the energy density in the magnetic field
at the shock between the pulsar wind and the Be-star wind
(Kirk, Ball \& Skj\ae raasen 1999), in which case the 
inverse Compton and synchrotron loss rates would be comparable.
However, the density of target photons available to
radio-emitting electrons originating in the disk is likely
to be low because of screening by the disk (Tavani \& Arons 1997),
and so inverse Compton losses are likely to be less important than
synchrotron losses in the model for the radio emission proposed here.

\begin{figure}[htb]
\centerline{\psfig{figure=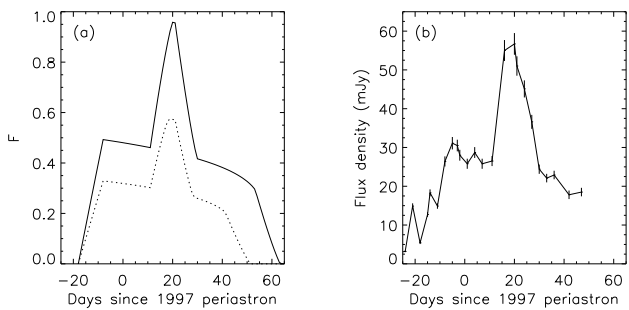,width=3.25in}}
\caption{(a) Model flux density $F(\nu)$ in arbitrary units
as a function of time (solid line),
and at a frequency $\nu'=1.44\,\nu$ (dashed line).
(b) Observed flux density from \PSR\ at 1.4~GHz
(from Figure 4 of Johnston et al.\ 1999).
}
\end{figure}

Assuming that the synchrotron emissivity can be approximated
by a delta-function in frequency,
the observed emission at a given frequency $\nu$
originates from electrons with characteristic
energies $\varepsilon_1$ 
and $\varepsilon_2$ at the two acceleration sites.
Since $\nu\propto B_{1,2}\varepsilon_{1,2}^2$ it follows that
in general $\varepsilon_1\neq \varepsilon_2$.
The total number density per unit energy
of electrons emitting characteristically at $\nu$
is then $N_1(\varepsilon_1,t)+N_2(\varepsilon_2,t)$.
When adiabatic losses (and therefore expansion) are neglected
it follows from (\ref{eq:synch}) that the 
that the emitted flux density is
proportional to the number density of electrons radiating at $\nu$.
The solid line in Figure 1a is the model flux density $F(\nu)$
obtained in this way, plotted in arbitrary units as a function of time.
The curve is obtained
by solving equation (\ref{eq:diffeqn}) 
for $T_1=T_2=10\,{\rm d}$,
$1/(b_1\varepsilon_1)=72\,{\rm d}$,
$1/(b_2\varepsilon_2)=9\,{\rm d}$, 
and source functions of the form
$Q_{1,2}(\varepsilon,t)\propto(\varepsilon/\varepsilon_0)^{-a}$
($\varepsilon_{min}\leq\varepsilon\ll\varepsilon_{max}$,
$a=2.2$).
The dashed curve shows the model flux density for
a frequency $\nu'=1.44\,\nu$.
Since the pulsar is closer to the Be star during the
post-periastron disk crossing than during the pre-periastron crossing, 
ram pressure balance implies $L_2<L_1$;
however, the choice $L_1/L_2 =4$ $(=B_2/B_1)$
in Figure 1a is purely illustrative because we have no precise
knowledge of the geometry of the pulsar-disk interaction.
The synchrotron-loss times 
$1/(b_{1,2}\varepsilon_{1,2})$ are roughly
the values expected for $B\sim 10^{-4}\,{\rm T}$ and
$\nu\sim 1\,{\rm GHz}$;
the injection intervals $T_{1,2}$ are roughly twice
the pulsar disk crossing times; the relative magnitudes
of $Q_1$ and $Q_2$ are chosen arbitrarily to be similar.
The curves are approximately piecewise-linear,
a characteristic of solutions to equation (\ref{eq:diffeqn})
for $a \approx 2$.
The proposed model reproduces many of the essential features of the
light curves in Figure 4 of Johnston et al.\ (1999),
from which the 1.4~GHz light curve is reproduced in Figure 1b.
In particular, the flux density is higher after $\peri+12$
because it is the sum of the contributions from two
accelerated electron populations.

The following features of the light curves in
Figures 1a and 1b should be noted.
(i) The first rise is associated with the pre-periastron
disk crossing and lasts for a time $T_1$, because the
synchrotron-loss time $1/(b_1\varepsilon_1)$
is much greater than $T_1$.
(ii) The gradual decay after $\peri-8$,
which continues until $\peri+54$ (solid curve), 
occurs because the number of electrons per unit time
cooling down to energy $\varepsilon_1$ from 
$\varepsilon > \varepsilon_1$
is slightly smaller than the number
with energy $\varepsilon_1$ cooling to lower
energies. (Exact balance occurs for $a=2$.)
(iii) The second rise is associated with the
post-periastron disk crossing and lasts for the
synchrotron-loss time $1/(b_2\varepsilon_2)$,
because $T_2$ is larger than $1/(b_2\varepsilon_2)$.
(iv) The plateau at the principal maximum
lasts for a time $T_2-1/(b_2\varepsilon_2) = 1\,{\rm d}$,
which is brief because the injection and loss times
for the post-periastron disk crossing
have been chosen to be almost equal. [cf.\ (ii)]
(v) Following this brief plateau,
$N_2(\varepsilon_2,t)$ decreases to zero over a time 
$1/(b_2\varepsilon_2)$ equal to the rise time in (iii).
Since $N_2(\varepsilon_2,t)$ decreases to zero
relatively rapidly,
its contribution drops below that of the more slowly
decaying $N_1(\varepsilon_1,t)$
and the subsequent evolution
is therefore an extension of (ii).
Note that the pre-periastron rise time (i) is determined by injection
(which lasts longer than the loss time), whereas the 
post-periastron  rise time (iii) is determined by the loss time
(which is shorter than the second injection episode).
These rise times happen to be similar because we chose
parameters such that $T_1\sim 1/(b_2\varepsilon_2)$. 

The changing separation of the solid and dashed curves in Figure 1a
indicates that the effective spectral index $\overline{a}$
of the composite electron distribution is not constant.
By crudely relating $\overline{a}$
to $\alpha$ ($S\propto \nu^{-\alpha}$) via 
$\alpha=(\overline{a}-1)/2$, we predict that the spectrum:
(i) has $\alpha\sim 0.7$ and does not change significantly from
switch-on at $\peri-18$ until the short plateau around $\peri+20$;
(ii) then steepens markedly with $\alpha$ increasing to $\sim 1.1$
before flattening again to $\sim 0.7$ at $\peri+30$;
(iii) steepens smoothly thereafter as the emission at 
higher frequencies decays more rapidly.
The model behavior is broadly in accord with the data,
which reveal a systematic steepening of the spectrum
with the spectral index changing by $\sim0.4$ between $\peri+15$
and $\peri+30$ (Figure 5 of Johnston et al.\ 1999),
followed by a less pronounced flattening.

The model makes a clear, testable prediction.
We see from Figure 1a that
the slow decay after the first disk crossing
(which continues well past the second disk crossing to the end of
the observation period $\peri+47$) ends sharply;
the model flux density decreases to zero over a 
time comparable to the initial rise, $T_1=10\,{\rm d}$,
instead of maintaining its very gradual downward slope 
all the way to zero. 
Although observations extending to $\peri+100$ were made at 0.84~GHz
following the 1994 periastron,
these single-frequency data are inconclusive as to the
existence of such a feature.
It will be possible to test this prediction 
during the next periastron passage; 
the frequency dependence of the onset of this rapid decline 
should also be observable.

The model makes no attempt to explain the shortest time scale
variations in the radio emission.
Particularly puzzling is the flat spectrum `blip' which was observed
at $\peri-21$ in 1997.
We speculate that this feature is the result of the pulsar
nebula splashing into the Be-star disk.
Such effects are more likely to have observable consequences the
first time the pulsar enters the disk because the system
geometry is such 
that we have an unobscured view of the pre-periastron
impact (see Figure 6 of Johnston et al.\ 1999), but
not of the post-periastron impact.

\section{Conclusions}

We have argued that the transient,
unpulsed radio emission from the binary system \PSR\
is synchrotron emission associated with the shock
interaction of the relativistic
pulsar wind and the dense circumstellar disk of the Be star.
The radiating electrons originate from the disk,
not a spherically-symmetric pulsar wind with the parameters
required to model the X-ray data.

We have presented a simple physical model for the evolution
of the unpulsed radio emission, based on impulsive
acceleration of electrons followed by synchrotron losses,
which reproduces the salient features of the radio light
curves and spectra observed during the
1994 and 1997 periastron passages.
The model also makes a testable prediction of a rapid 
frequency-dependent decay in the radio flux
several weeks after the principal flux density peak,
to be searched for during
observations of the next periastron passage in October 2000.

\acknowledgments

O.S. would like to thank the Norwegian Research Council
for support of this work through a postgraduate grant,
and the RCfTA, University of Sydney for general support.
A.M. gratefully acknowledges financial support from
the Miller Institute for Basic Research in Science through
a Miller Fellowship.

\end{document}